\documentclass[aps,prd,amsmath,amssymb,twocolumn,nofootinbib,floatfix]{revtex4-2}

\usepackage{graphicx}
\usepackage{dcolumn}
\usepackage{bm}
\usepackage{multirow}
\usepackage{booktabs}
\usepackage[normalem]{ulem}

\usepackage{siunitx}
\usepackage{xcolor}
\usepackage{hyperref}

\definecolor{citecolor}{rgb}{0.016, 0.565, 0.929}
\definecolor{nicered}{rgb}{0.7,0.1,0.1}
\definecolor{nicegreen}{rgb}{0.1,0.5,0.1}
\hypersetup{colorlinks,citecolor= nicered,linkcolor= blue,urlcolor=citecolor}

\begin{document}

\title{Updated and Projected Cosmic Microwave Background Bounds on WIMP Annihilation}

\author{Charlotte Myers}
\affiliation{MIT Department of Physics, 77 Massachusetts Ave., Cambridge, MA 02139, USA}
\author{Dominic Agius}
\affiliation{Instituto de Fisica Corpuscular (IFIC),  CSIC‐Universitat de València, 46071, Valencia, Spain}
\author{Daniele Gaggero}

\affiliation{INFN Sezione di Pisa, Polo Fibonacci, Largo B. Pontecorvo 3, 56127 Pisa, Italy}
\author{Angelo Ricciardone}
\affiliation{Dipartimento di Fisica ``Enrico Fermi'', Università di Pisa,
Largo Bruno Pontecorvo 3, Pisa I-56127, Italy}
\affiliation{INFN Sezione di Pisa, Polo Fibonacci, Largo B. Pontecorvo 3, 56127 Pisa, Italy}

\date{\today}

\begin{abstract}
We derive updated Cosmic Microwave Background (CMB) constraints on annihilating dark matter, and present forecasts for upcoming CMB surveys. We show that the addition of recent temperature, polarization, and lensing data from ground--based experiments yields only minor improvements ($\approx 10\%$) compared to Planck bounds, confirming that the sensitivity remains dominated by the large-scale E-mode polarization. Forecasts, using a LiteBIRD-like setup, indicate that pairing a low-noise, wide-sky satellite at $\ell < 200$ with high-resolution ground observations nearly saturates the cosmic-variance limit, improving bounds by $\approx 60\%$, where our derived 95th percentile limit is $p_{\rm ann} < 1.27{\times}10^{-28}\,\mathrm{cm^{3}\,s^{-1}\,GeV^{-1}}$. We also consider the inclusion of B-mode polarization for a realistic future experiment.

\end{abstract}

\maketitle

\section{\label{sec:level1}Introduction }
The Weakly Interacting Massive Particle (WIMP) paradigm remains one of the most widely studied and theoretically well-motivated dark matter (DM) scenarios~\cite{Arcadi:2017kky}. Within this framework, the generic prediction of DM annihilation into Standard Model particles implies an energy injection that alters the thermal and ionization history of the universe. This impact is measurable during the cosmic dark ages---the period between recombination and reionization---when the intergalactic medium is otherwise nearly neutral and astrophysical sources are absent. Through an exotic electromagnetic injection, annihilating DM leaves a distinct imprint on the cosmic microwave background (CMB) by increasing the free-electron abundance and enhancing Thomson scattering with CMB photons, thereby modifying the Thomson optical depth. This indirect observation channel has been studied in detail, leading to some of the most stringent constraints on DM annihilation~\cite{Adams:1998nr,Padmanabhan:2005es,Galli:2009zc,Slatyer:2009yq,Kanzaki:2009hf,Hutsi:2011vx,Galli:2011rz,Finkbeiner:2011dx,Giesen:2012rp,Cline:2013fm,Weniger:2013hja,Lopez-Honorez:2013cua,Diamanti:2013bia,Madhavacheril:2013cna,Galli:2013dna,Planck:2015fie,Slatyer:2015jla,Slatyer:2015kla,Slatyer:2016qyl,Kawasaki:2021etm,Cirelli:2024ssz}.

With the advent of precision cosmology, it became possible to constrain the DM annihilation cross section using CMB polarization data~\cite{Padmanabhan:2005es}. This effort  built on earlier works examining the impact of decaying DM on the recombination history and the resulting CMB signatures~\cite{Peebles:2000pn, Bean:2003fb, Doroshkevich:2002wy, Chen:2003gz, Hansen:2003yj, Pierpaoli:2003rz}. Subsequent studies refined these analyses by developing increasingly accurate treatments of energy injection and deposition during the dark ages~\cite{Galli:2009zc,Slatyer:2009yq}. In these studies, it was recognized that the bulk of the constraining power, when considering changes to the Thomson optical depth induced from additional ionizations from DM annihilation, comes from large angular scales, with sensitivity dominated by the low–multipole polarization~\cite{Padmanabhan:2005es,Galli:2010it,Madhavacheril:2013cna}. This emphasis on low–$\ell$ $E$-mode polarization, and the corresponding insensitivity of small-scale TT data, once parameter degeneracies are marginalized, was sharpened in later analyses, which also clarified why polarization systematically outperforms temperature measurements for energy injection constraints~\cite{Galli:2014kla}. In parallel, several forecast studies compared the capability of Planck with those of cosmic-variance-limited (CVL) and ground-based experiments~\cite{Galli:2010it,Wu:2014hta,Calabrese:2016eii}. Subsequent work further identified the redshift ranges and multipole scales most relevant for DM–induced energy deposition, providing a unified framework for understanding how the deposition history maps onto CMB observables~\cite{Green:2018pmd}. 

Amid this progress, forecasts for next generation high-resolution, ground-based CMB experiments, such as CMB-S4~\cite{CMB-S4:2016ple}, indicated substantial improvements over Planck in their ability to constrain DM energy injection~\cite{Madhavacheril:2013cna,Wu:2014hta,Cang:2020exa}. While Ref.~\cite{Madhavacheril:2013cna} anticipated moderate improvements from CMB-S4 over Planck, Refs.~\cite{Wu:2014hta,Cang:2020exa} initially claimed an order-of-magnitude improvement. However, the latter estimate was later revised, indicating a more modest enhancement of only a factor of $\sim2$~\cite{Zhang:2023usm}. This reinforced the general consensus that, once low-$\ell$ large-scale polarization is constrained, additional high-$\ell$ sensitivity yields only marginal improvements. Recent forecasts including a LiteBIRD-like satellite reinforce this picture: combining space-based, low-noise $E$-mode polarization at $\ell\lesssim100$ with high-resolution ground-based data effectively saturates the information available from the CMB and approaches the cosmic-variance limit for realistic sky coverage~\cite{Wang:2025tdx}. In this work, we revisit current bounds in the post-Planck era using updated ACT, SPT, and DESI measurements. We also present forecasts that explicitly quantify where the information resides in multipole space and how close future survey combinations can come to the CVL floor. Moreover, we also consider the phenomenological imprint on $B$-mode polarization data. 

The rest of this paper is organized as follows. In Sec.~\ref{sec:Constraining Power} we review how DM annihilation modifies the ionization history and CMB anisotropies, introduce the parameter $p_{\mathrm{ann}}$, and quantify the multipole ranges that drive the constraining power using Fisher forecasts. In Sec.~\ref{sec:Updated Bounds} we update current limits on $p_{\mathrm{ann}}$ by combining Planck with recent ACT and SPT data, as well as BAO data, and examine the resulting parameter degeneracies. Sec.~\ref{sec:Projected Bounds} presents forecasts for future CMB configurations, including LiteBIRD- and CMB-S4–like surveys and their combinations, using both Fisher analyses and MCMC on mock data, and maps these constraints into the $(m_\chi,\langle\sigma v\rangle)$ plane for representative annihilation channels. In Sec.~\ref{sec: Effect of Tensor Modes} we discuss the information carried by $B$-mode polarization and quantify its impact on bounds relative to $E$-mode–dominated constraints. We summarize our main conclusions and discuss implications for future CMB polarization efforts in Sec.~\ref{sec:Discussion}, with priors, numerical details, and full posterior distributions included in the appendices.

\section{Constraining Power} \label{sec:Constraining Power}

CMB constraints on DM annihilation are conveniently expressed in terms of the model-independent parameter $p_{\mathrm{ann}}$~\cite{Galli:2009zc}:
\begin{equation}\label{eqn:p_ann}
    p_{\mathrm{ann}} \equiv f_{\mathrm{eff}} \frac{\langle \sigma v \rangle}{m_\chi} \,,
\end{equation}
where $f_{\mathrm{eff}}$ encodes the fraction of injected energy that is efficiently deposited into the medium, $\langle \sigma v \rangle$ is the thermally averaged cross section, and $m_\chi$ is the DM particle mass~\cite{Slatyer:2009yq, Madhavacheril:2013cna}. For WIMP annihilation, the CMB is primarily sensitive to the overall amplitude of the energy deposition curve rather than its detailed redshift dependence. As a result, a single parameter $f_{\mathrm{eff}}$ accurately captures the impact of a given final state, allowing bounds on $p_{\mathrm{ann}}$ to be easily translated to any annihilation channel of interest~\cite{Slatyer:2015jla}.

\begin{figure}
    \centering
    \includegraphics[width=1\linewidth]{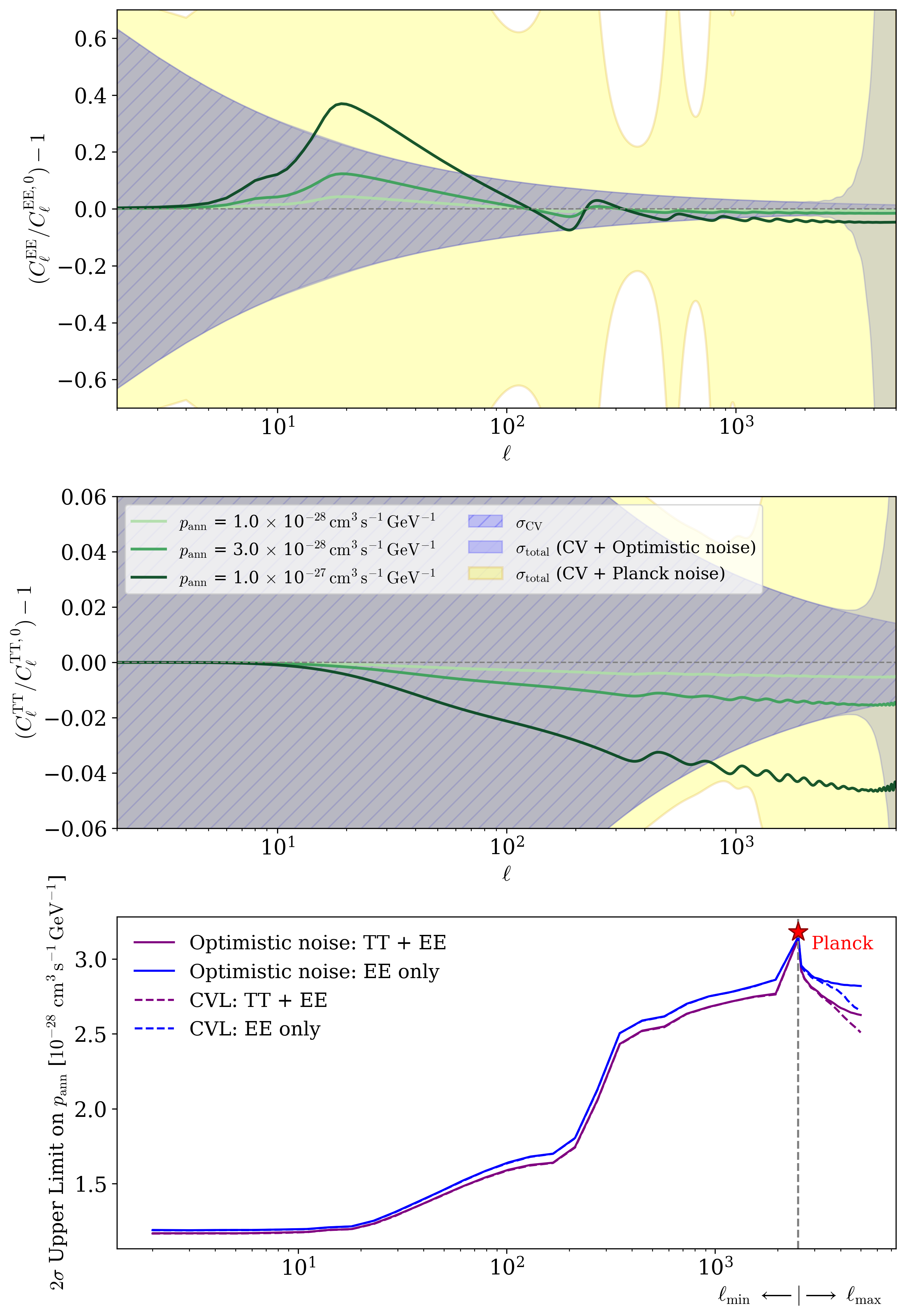}
    \caption{
    \textbf{Top:} Fractional impact of varying $p_{\mathrm{ann}}$ on the EE and TT spectra. Outer shaded bands indicate cosmic variance + instrumental noise, and inner shaded bands indicate cosmic variance alone; both are computed for the highest value of $p_{\mathrm{ann}}$ ($1.0 \times 10^{-27}\,\mathrm{cm}^3\,\mathrm{s}^{-1}\,\mathrm{GeV}^{-1}$).
    \textbf{Middle:} Same as middle panel but for the TT power spectrum.
    \textbf{Bottom:} Forecasted $2\sigma$ upper limits on $p_{\mathrm{ann}}$ for different experimental configurations and multipole ranges included in the Fisher analysis.
    Solid curves include TT{+}EE; dashed curves include EE-only. A Planck-like experiment with $\ell_\text{max} = 2500$ is taken as a baseline (marked by star).}
    \label{fig:3panel_forecast}
\end{figure}

DM annihilation injects energetic particles that heat and ionize the neutral intergalactic medium (IGM) \cite{Slatyer:2015jla}. The dominant impact on the CMB is an increase in the free–electron fraction $x_e(z)$, which broadens the visibility function and adds an effective optical depth $\Delta\tau_{\rm inj}$ between recombination and reionization. To leading order, this multiplies the small–scale anisotropies by \cite{Natarajan:2012ry, Green:2018pmd}
\begin{equation}
C_\ell^{TT,EE}\;\rightarrow\;e^{-2\Delta\tau_{\rm inj}}\,C_\ell^{TT,EE}\,,
\end{equation}
producing a damping of the TT/EE tails that is strongly degenerate with shifts in amplitude $A_s$ and tilt $n_s$ of the scalar power spectrum. This suppression is evident in the top and middle panels of Fig.~\ref{fig:3panel_forecast}.

Annihilation also enhances large–scale polarization by generating additional $E$-modes through Thomson scattering, yielding a broad bump in the EE spectrum for $\ell\!\lesssim\!100$ (top panel) \cite{Slatyer:2009yq, Green:2018pmd, Padmanabhan:2005es}. Unlike a pure change in $\tau_{\rm reio}$, this feature reflects energy deposition over a wide redshift range and has a distinct $\ell$-dependence, which helps to break parameter degeneracies \cite{Green:2018pmd}.

To localize where the sensitivity to $p_{\mathrm{ann}}$ originates, we compute the Fisher information as a function of the multipole range under two benchmark survey models: an “Optimistic noise’’ instrument (i.e., white noise, $1'$ FWHM, $\Delta_T=1~\mu\mathrm{K}\,\mathrm{arcmin}$ with $\Delta_P=\sqrt{2}\,\Delta_T$, $f_{\mathrm{sky}}=0.7$) and, for reference, a CVL survey with the same $f_{\mathrm{sky}}$. In the bottom panel of Fig.~\ref{fig:3panel_forecast} we vary a transition multipole $\ell_{\min}$ that controls a hybrid noise model: for $\ell<\ell_{\min}$ we replace the Planck noise with the Optimistic (or CVL) noise, while for $\ell\ge \ell_{\min}$ we retain Planck noise levels. Thus moving $\ell_{\min}$ to smaller values progressively upgrades more large-scale modes. The resulting forecasted $2\sigma$ upper limit on $p_{\mathrm{ann}}$ tightens rapidly as $\ell_{\min}$ drops from $\sim300$ to $\sim20$, with continued (but slower) improvement down to reionization scales. The near coincidence of the TT{+}EE and EE-only curves shows that, after marginalizing parameter degeneracies, the gain comes almost entirely from large-scale $E$-modes; by contrast, extending to higher $\ell$ yields only marginal additional benefit beyond Planck.

\section{Updated Bounds} \label{sec:Updated Bounds}

While Planck provided a nearly full-sky measurement of the CMB temperature and polarization anisotropies with unprecedented precision on large angular scales \cite{Planck:2018vyg}, recent ground-based CMB experiments have extended sensitivity to  smaller scales. The Atacama Cosmology Telescope
(ACT), located in the Atacama Desert, has delivered the most precise measurements of the temperature power spectrum at multipoles $\ell > 1800$, and of the E-mode polarization spectrum for $\ell > 750$ \cite{ACT:2025xdm,ACT:2025fju,ACT:2025tim}. The South Pole Telescope (SPT)-3G, located at the South Pole, offers sensitivity competitive with ACT between $\ell = [1800, 4000]$ for the EE spectrum, and between $\ell = [2200, 4000]$ for the TE spectrum \cite{SPT-3G:2025bzu}.

Since DM annihilation has an effect on the damping tail of both the TT and EE spectra (though to a smaller extent compared to large-scale polarization), we reassess current bounds in the post-Planck era by combining Planck with ground-based measurements that provide enhanced high-$\ell$ sensitivity.

\subsection{Datasets}

As Planck remains the most sensitive experiment at large angular scales, we combine recent datasets with Planck to maximize their joint sensitivity.  Details of these combinations are given in Table~\ref{tab:dm_ann_limits}. 

In adding ACT, we use the P-ACT combination as described in \cite{ACT:2025tim}, which includes multipole cuts designed to optimize the joint constraining power of Planck and ACT while minimizing overlap. In combining with SPT, we assume that both Planck and ACT are sufficiently uncorrelated with SPT to justify adding SPT without any multipole range cuts, justified by the minimal sky overlap between SPT and other experiments \cite{SPT-3G:2025bzu}. We additionally assume the ACT and SPT lensing data are sufficiently uncorrelated to be directly combined \cite{SPT-3G:2025bzu}. 

We also consider the effect of including recent DESI DR2 baryon acoustic oscillation (BAO) measurements. Supernova (SN) data are omitted, as they constrain late-time expansion rather than the early-Universe physics relevant for energy injection constraints~\cite{2024IAUGA..32P2915R_sne, Madhavacheril:2013cna,ACT:2025tim}. 

\subsection{Methodology}  \label{sec:bounds_methodology}

We use the Boltzmann solver code \texttt{CLASS} \cite{Blas:2011rf} to compute theoretical power spectra, and the Markov Chain Monte Carlo (MCMC) sampler \texttt{Cobaya} to explore the posterior distribution of the cosmological parameters \cite{ Torrado:2020dgo,Cobaya_ASCL:2019}. In our analysis, we vary the six standard $\Lambda$CDM parameters ($\{A_s, n_s, \theta_s, \tau_{reio}, \Omega_b h^2, \Omega_c h^2\}$), with fiducial values fixed by Planck \cite{Planck:2018vyg}), and include the annihilation parameter $p_{\mathrm{ann}}$ as an additional degree of freedom.

The impact of DM annihilation is modeled as an additional energy injection term that affects the recombination history, implemented through the built-in parameterization of energy deposition in \texttt{CLASS} \cite{Blas:2011rf}, built on {\tt ExoCLASS}~\cite{Stocker:2018avm}.

To ensure convergence, we run multiple independent chains and require a Gelman-Rubin convergence criterion of $R - 1 < 0.01$~\cite{gelman_inference_1992}.  All chains use a flat prior on $p_\text{ann} \geq 0 $, and flat priors on all cosmological parameters, provided in Table~\ref{tab:priors}. 


\subsection{Results}\label{sec:data_results}

Our results are reported in Table \ref{tab:dm_ann_limits}, with the triangle plot showing pairwise marginalized posteriors in the Appendix. In Section \ref{sec:Results}, we interpret the most stringent results for relevant annihilation channels and compare them to existing astronomical limits. We find that the most constraining combination is Planck + SPT; the addition of SPT tightens the bound by approximately $10\%$, compared to Planck alone. The addition of ACT data slightly weakens the bound, and the addition of DESI DR2 BAO data produces no statistically significant improvement. This is explained by a marginal shift away from zero in the $p_{\rm ann}$ posterior peak when combining ACT and Planck data (see, e.g., Fig.~33 of~\cite{ACT:2025tim}, and the discussion therein). 

We obtain a slightly weaker bound for Planck-only than the official Planck 2018 result. We attribute this to our use of the \texttt{SRoll2} low-$\ell$ likelihood, which incorporates improved systematics and map-making. When we substitute the original Planck low-$\ell$ likelihood, we recover the 2018 bound of $3.2 \times 10^{-28}\,\mathrm{cm}^3\,\mathrm{s}^{-1}\,\mathrm{GeV}^{-1}$, a difference of $\sim15\%$. This difference highlights the significant role of low-$\ell$ polarization data, where improved systematics have a dominant effect on the $p_{\text{ann}}$ constraint compared to the addition of high-$\ell$ data.

Our results are consistent with previous analyses that indicate that most of the available constraining power will come from improved measurements of large-scale polarization \cite{Padmanabhan:2005es, Green:2018pmd, Slatyer:2015jla}.

\begin{table*}
\caption{\label{tab:dm_ann_limits} $2 \sigma$ upper limits on $p_{\mathrm{ann}}$ for different experimental datasets}
\begin{tabular}{l @{\hspace{1cm}} p{5cm} c}
\toprule
\textbf{Configuration} & \textbf{Datasets Included} & \textbf{2$\sigma$ Upper Limit  [$10^{-28}\,\mathrm{cm}^3\,\mathrm{s}^{-1}\,\mathrm{GeV}^{-1}$]} \\
\midrule
Planck Only &
Planck 2018 low-$\ell$ TT \newline
Planck 2018 low-$\ell$ EE (SRoll2) \newline
Planck 2018 high-$\ell$ TTTEEE \newline
Planck PR4 (2020) NPIPE Lensing &
3.9  \\
\\[-1.5ex]
Planck + SPT &
\textit{Planck Only} \newline
SPT-3G D1 Lite (TnE) \newline
MUSE lensing&
3.38  \\
\\[-1.5ex]
Planck + ACT &
\textit{P-ACT} \newline
ACT-DR6 Lenslike  &
4.13  \\
\\[-1.5ex]
Planck + ACT + SPT &
\textit{P-ACT} \newline
SPT-3G D1 Lite (TnE) \newline
MUSE lensing \newline
ACT-DR6 Lenslike &
3.73 \\
\\[-1.5ex]
Planck + ACT + SPT + DESI &
\textit{Planck + ACT + SPT} \newline
DESI DR2 BAO &
3.70 \\
\bottomrule
\end{tabular}
\end{table*}

\section{Projected Bounds} \label{sec:Projected Bounds}

Future satellite missions will surpass Planck in sensitivity to large-scale polarization, providing full-sky, low-noise observations at the multipoles most sensitive to energy injection. Ground-based experiments such as the Simons Observatory will contribute complementary small-scale measurements, but remain limited due to low sky coverage and poor sensitivity to low-$\ell$ modes compared to satellite missions. We present forecasts on $p_{\mathrm{ann}}$ considering multiple experimental scenarios, comparing results from Fisher-matrix calculations and MCMC analyses.

\subsection{Future CMB Surveys}

A new generation of CMB experiments will improve upon Planck’s sensitivity both through better angular resolution and through lower noise across a wide range of multipoles:

\paragraph{LiteBIRD.}
LiteBIRD is a JAXA-led satellite mission scheduled for launch in the 2030s, with a primary science goal of detecting primordial gravitational waves via large-scale B-mode polarization \cite{LiteBIRD:2022cnt}. The large sky coverage ($f_{\mathrm{sky}} \gtrsim 0.7$) and strong control over systematics make LiteBIRD especially valuable for low-$\ell$ polarization measurements, which are highly sensitive to changes in the ionization history from DM annihilation. In contrast to ground-based experiments, LiteBIRD’s space-based platform avoids atmospheric noise and allows access to the reionization bump at $\ell \lesssim 10$, where Planck noise remains a limiting factor \cite{LiteBIRD:2022cnt, Sailer:2021yzm}.

\paragraph{CMB-S4 (concept).}
Although the CMB-S4 program has been largely defunded, we treat it here as a representative high-resolution ground-based experiment for comparative purposes \cite{CMB-S4:2016ple,CMBS4Web}. The CMB-S4 concept called for an array of telescopes at the South Pole and in the Atacama Desert, with combined sensitivities approaching $\sim 1~\mu\mathrm{K}$-arcmin in temperature \cite{Euclid:2021qvm}. For DM annihilation, such high-resolution data primarily improve constraints via the damping tail of the CMB power spectra ($\ell \gtrsim 1000$). 

\paragraph{Simons Observatory.}
The Simons Observatory (SO), currently operating in the Atacama Desert, will serve as the next major high-resolution ground-based CMB experiment. SO involves a combination of large- and small-aperture telescopes to target both large- and small-scale anisotropies, with a particular emphasis on high-$\ell$ measurements of temperature and E-mode polarization \cite{SimonsObservatory:2018koc, SO:2024ntl}.

\subsection{Methodology}
\subsubsection{Noise Modeling}

We model each frequency channel $i$ with a Gaussian beam of full width at half maximum
$\theta^{\mathrm{FWHM}}_i$ (in radians) and white noise levels $\Delta_{T,i}$ and $\Delta_{P,i}$
(in $\mu\mathrm{K}$-arcmin). The corresponding beam dispersion and window function are given as \cite{Knox:1995dq}:
\begin{align}
\sigma_i &= \frac{\theta^{\mathrm{FWHM}}_i}{\sqrt{8\ln 2}}, \\
b_{\ell,i} &= \exp\!\left[-\tfrac{1}{2}\,\ell(\ell+1)\,\sigma_i^2\right].
\end{align}
Map noise in $\mu\mathrm{K}$-arcmin is converted to $\mu\mathrm{K}$-radian via
\begin{equation}
\sigma_{X,i} \;=\; \Delta_{X,i}\times \left(\frac{\pi}{180\times 60}\right), \qquad X\in\{T,\,P\}.
\end{equation}
Assuming uncorrelated channels that are beam-deconvolved at the power-spectrum level,
the effective noise power spectra are obtained by inverse-variance combining channels:
\begin{align}
N_\ell^{TT} &= \left[ \sum_i \sigma_{T,i}^{-2}\, b_{\ell,i}^{\,2} \right]^{-1}, \\
N_\ell^{EE} &= \left[ \sum_i \sigma_{P,i}^{-2}\, b_{\ell,i}^{\,2} \right]^{-1}, \\
N_\ell^{TE} &= 0,
\end{align}
where $N_\ell^{TE}=0$ reflects the assumption of uncorrelated temperature and polarization noise.
Unless experiment-specific polarization sensitivities are provided, we adopt the conventional
$\Delta_{P,i}=\sqrt{2}\,\Delta_{T,i}$.

We provide details of the noise models for each experiment in Table~\ref{tab:forecast_specs}. 
For all experiments, we assume efficient removal of foregrounds, and note that the effect of foreground removal method has a minor but well-understood effect on the bound~\cite{Zhang:2023usm}.

Due to the contamination by atmospheric fluctuations and instrumental drifts, which typically scales as $1/f$, low-$\ell$ noise must be modeled carefully. For modeling ground-based experiments, we replace low-$\ell$ measurements ($\ell <$ 50) with Planck noise. To accurately represent the systematics of Planck, we include a standard boost factor of 8 on low-$\ell$ polarization noise \cite{Euclid:2021qvm, Planck:2018vyg}. We validate that this factor reproduces the true Planck bounds to within $10\%$. 

We assume LiteBIRD’s rotating half-wave plates (HWP) suppress $1/f$ noise sufficiently to include modes down to $\ell = 2$ \cite{Micheli:2024hfe, LiteBIRD:2022cnt}.

Due to the dominant effect of foregrounds on the temperature spectrum for $\ell > 3000$, we include only polarization measurements in the range $\ell = [3000, 5000]$ for forecasts on ground-based experiments \cite{Wu:2014hta}.

\subsubsection{Fisher Forecasting}\label{sec:fisher}

 Fisher information provides a convenient metric to estimate the precision with which model parameters can be constrained. The associated covariance matrix is given as \cite{Finkbeiner:2011dx}: 
\begin{equation} \label{eq:fisher_matrix_base}
\Sigma_\ell = \frac{2}{(2\ell + 1) f_{\mathrm{sky}}} \times 
\begin{pmatrix}
C_\ell^{TT} + N_\ell^{TT} & C_\ell^{TE} \\
C_\ell^{TE} & C_\ell^{EE} + N_\ell^{EE}
\end{pmatrix}
\end{equation}

Temperature and polarization measurements are assumed to be independent such that the noise of the TE cross spectrum vanishes. For a cosmic variance-limited experiment,  $N_\ell^{TT} = N_\ell^{EE} = 0$. The Fisher information can then be computed as:

\begin{equation}
F_{ij} = \sum_{\ell} 
\left( 
\frac{\partial C_\ell}{\partial \alpha_i} 
\right)^{\!\!T}
\cdot \Sigma_\ell^{-1} \cdot 
\left( 
\frac{\partial C_\ell}{\partial \alpha_j} 
\right)
\end{equation}
 where $\alpha$ denotes the cosmological parameters with respect to which the derivatives are taken. 
The derivatives are computed via finite differences with step sizes chosen to optimize numerical stability.

\subsubsection{Forecasting via MCMC Sampling}\label{sec:mcmc}

To validate our Fisher forecasts, we estimate the sensitivity of future experiments using MCMC simulations drawn from Gaussian approximations to the likelihood.  The likelihood is evaluated on mock data (fiducial spectra plus noise), and we assume a Wishart-distributed covariance estimator for each multipole in the specified range. 

Let $\hat{C}_\ell^{XY}$ denote the observed (mock) power spectra and $C_\ell^{XY}$ the theoretical spectra, where $X, Y \in \{T, E\}$. The total covariance matrix at each multipole $\ell$ is given by:
\begin{equation}
\mathbf{C}_\ell^{\text{theory}} =
\begin{pmatrix}
C_\ell^{TT} + N_\ell^{TT} & C_\ell^{TE} \\
C_\ell^{TE} & C_\ell^{EE} + N_\ell^{EE}
\end{pmatrix},
\end{equation}
where $N_\ell^{TT}$ and $N_\ell^{EE}$ denote the noise power spectra. The corresponding matrix of mock data is given by:
\begin{equation}
\mathbf{\hat{C}}_\ell =
\begin{pmatrix}
\hat{C}_\ell^{TT} & \hat{C}_\ell^{TE} \\
\hat{C}_\ell^{TE} & \hat{C}_\ell^{EE}
\end{pmatrix}.
\end{equation}

The log-likelihood is then computed as:
\begin{equation}
\begin{split}
-2 \log \mathcal{L}
&= f_{\mathrm{sky}} \sum_{\ell=\ell_{\min}}^{\ell_{\max}} (2\ell + 1)
\Bigl[
\operatorname{Tr}\!\bigl(\hat{\mathbf C}_\ell \mathbf C_\ell^{-1}\bigr) \\
&\qquad - \log \det\!\bigl(\hat{\mathbf C}_\ell \mathbf C_\ell^{-1}\bigr)
- 2
\Bigr].
\end{split}
\end{equation}

We use \texttt{Cobaya} to sample from the mock likelihoods, using the same parameterization and convergence criteria described in Section \ref{sec:bounds_methodology}.

\begin{figure*}
    \centering
    \includegraphics[width=0.8\linewidth]{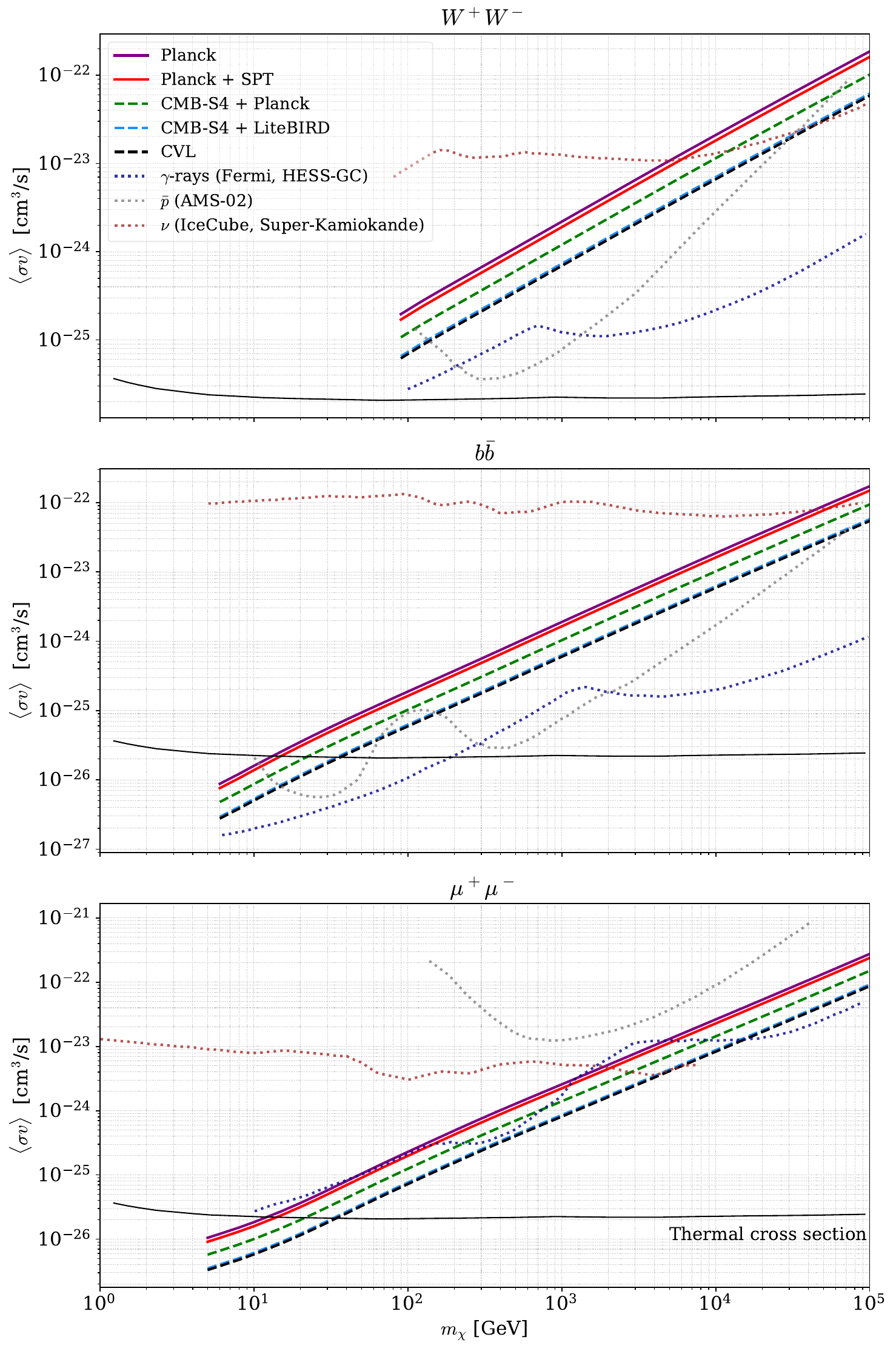}
    \caption{Projected $2\sigma$ upper bounds on the DM annihilation cross section $\langle\sigma v\rangle$ as a function of DM mass $m_\chi$ for three annihilation channels:  $W^+W^-$ (top), $b\bar{b}$ (middle), and $\mu^+\mu^-$ (bottom). The channel-dependent $f_{\mathrm{eff}}$ curves used to convert constraints on $p_{\mathrm{ann}}$ to $\langle\sigma v\rangle$ are taken from Ref.~\cite{Madhavacheril:2013cna}. 
    Solid lines show current CMB constraints derived in this work for Planck and Planck + SPT; dashed lines show projected bounds for CMB-S4 + Planck, CMB-S4 + LiteBIRD, and a CVL experiment. 
    Dotted lines indicate complementary limits from indirect detection, extrapolated from~\cite{Cirelli:2024ssz} (see also references therein). We also show the thermal cross section as a function of DM mass~\cite{Steigman:2012nb}.}
    \label{fig:mass_constraints}
\end{figure*}

\begin{table*}
\centering
\renewcommand{\arraystretch}{1.3}
\caption{\label{tab:forecast_specs}Experimental configurations (beam widths and noise levels per frequency channel) used in the forecasts. Where polarization noise is not provided, we adopt the standard $\Delta_{P}=\sqrt{2}\,\Delta_{T}$. }
\begin{tabular}{l c c c c p{5cm}}
\toprule
Experiment & $\nu$ [GHz] & Beam FWHM [arcmin] & $\Delta_T \ (\Delta_P)$ [$\mu$K-arcmin]  & $f_{\mathrm{sky}}$ & $\ell$-range \\
\midrule

\multirow{3}{*}{\textbf{Planck} \cite{Planck:2018vyg}} &
100 & 10.0 & 65 (103) & \multirow{3}{1cm}{0.7} & \multirow{3}{3cm}{2 $< \ell <$ 2500} \\
& 143 & 7.0 & 43 (81) & \\
& 217 & 5.0 & 66 (134) & \\

\midrule

\textbf{CMB-S4} \cite{Euclid:2021qvm, Wu:2014hta} & -- & 1.0 & 1.0 & \multirow{1}{1cm}{0.4} &
\shortstack[l]{T: $50<\ell<3000$\\ P: $50<\ell<5000$} \\

\midrule

\textbf{Simons Observatory} \cite{SimonsObservatory:2018koc} &  &  &  &  &  \\
\multirow{6}{*}{\textit{Small Aperture Telescopes}} &
27& 91 & 35  & \multirow{6}{1cm}{0.7} &  \multirow{6}{5cm}{$50 < \ell <$ 5000} \\
& 39 & 63 & 21 & \\
& 93 & 30 & 2.6  & \\
& 145 & 17 & 3.3  & \\
& 225 & 11 & 6.3 & \\
& 280 & 9 & 16 & \\

\vspace{0.5cm}
\multirow{6}{*}{\textit{Large Aperture Telescopes}} &
27& 7.4 & 71  & \multirow{6}{1cm}{0.2} &  \multirow{6}{5cm}{50 $< \ell <$ 5000} \\
& 39 & 5.1 & 36 & \\
& 93 & 2.2 & 8  & \\
& 145 & 1.4 & 10  & \\
& 225 & 1.0 & 22 & \\
& 280 & 0.9 & 53 & \\

\midrule

\multirow{15}{*}{\textbf{LiteBIRD} \cite{LiteBIRD:2022cnt}} &
40  & 70.5          & 26.46 & \multirow{15}{1cm}{0.7} & \multirow{15}{5cm}{$2 \le \ell \le 200$} \\
& 50  & 58.5          & 23.66 & \\
& 60  & 51.1          & 15.07 & \\
& 68  & 41.6, 47.1    & 14.08, 22.47 & \\
& 78  & 36.9, 43.8    & 11.00, 13.53 & \\
& 89  & 33.0, 41.5    & 8.68, 20.34  & \\
& 100 & 30.2, 37.8    & 7.31, 6.00   & \\
& 119 & 26.3, 33.6    & 5.44, 4.03   & \\
& 140 & 23.7, 30.8    & 5.13, 4.51   & \\
& 166 & 28.9          & 3.94         & \\
& 195 & 28.0, 28.6    & 4.99, 7.43   & \\
& 235 & 24.7          & 7.63         & \\
& 280 & 22.5          & 9.76         & \\
& 337 & 20.9          & 15.52        & \\
& 402 & 17.9          & 33.55        & \\

\midrule

\bottomrule
\end{tabular}
\end{table*}

\begin{figure}[t]
    \centering
    \includegraphics[width=1.0\linewidth]{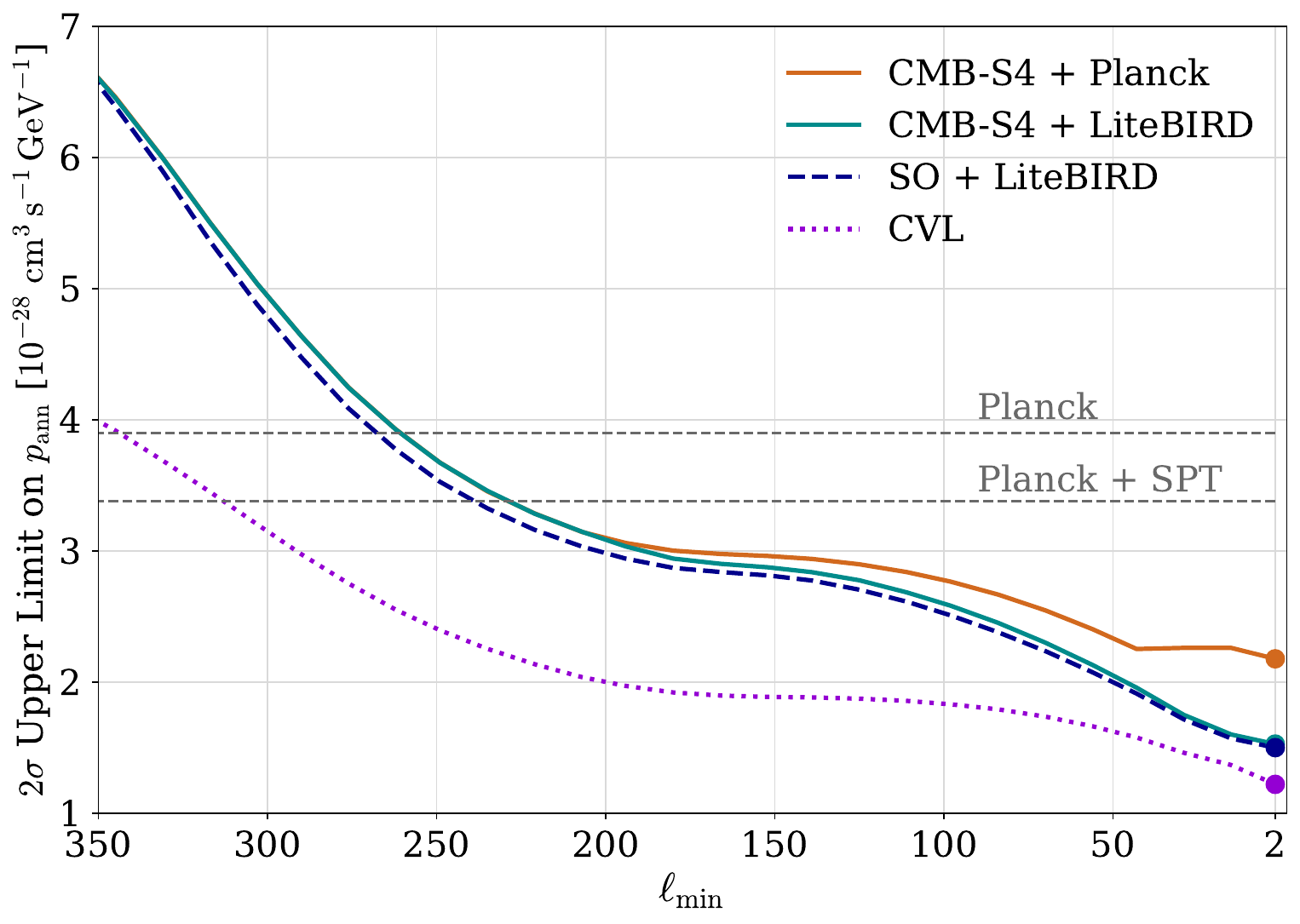}
    \caption{\label{fig:bounds}
    Fisher-forecasted $2\sigma$ upper limit on $p_{\mathrm{ann}}$ as a function of the minimum multipole $\ell_{\min}$ included in the analysis. For $\ell > 200$, forecasts assume either a CMB-S4–like experiment (solid lines), or a Simons Observatory–like configuration for \textit{Simons + LiteBIRD} (purple dashed); see Table \ref{tab:forecast_specs} for configuration details. For $\ell < 200$, each forecast uses the low-$\ell$ experiment indicated in the legend. The CVL forecast (dotted) assumes coverage of $f_{\mathrm{sky}} = 0.7$ and sensitivity over the entire multipole range up to $\ell = 5000$, without  imposing any $\ell_{\min}$ cut.
    Horizontal dashed lines refer to bounds computed in this work (see Table \ref{tab:dm_ann_limits}).}
    
    \label{fig:lmin_exps}
\end{figure}

\subsection{Dataset Combinations}

We evaluate several forecast configurations that combine different experiments across complementary multipole ranges, chosen to maximize joint sensitivity (see Table~\ref{tab:forecast_specs}). For each configuration, we construct hybrid spectra by selecting the experiment with the best expected performance in the relevant $\ell$ range: 

\begin{itemize}
    \item \textbf{CMB-S4 + Planck}:
    \begin{itemize}
        \item $\ell < 50$: Planck
        \item $\ell \in [50, 3000]$: CMB-S4 TT,TE,EE
        \item $\ell\in [3000, 5000]$: CMB-S4 EE-only 
    \end{itemize}

    \item \textbf{CMB-S4 + LiteBIRD}:
    \begin{itemize}
        \item $\ell < 200$: LiteBIRD
        \item $\ell \in [200, 3000]$: CMB-S4 TT,TE,EE
        \item $\ell\in [3000, 5000]$: CMB-S4 EE-only 
    \end{itemize}

    \item \textbf{SO + LiteBIRD}:
    \begin{itemize}
        \item $\ell < 200$: LiteBIRD
        \item $\ell \in [200, 3000]$: SO TT,TE,EE
        \item $\ell\in [3000, 5000]$: SO EE-only 
    \end{itemize}

    \item \textbf{CVL}:
    \begin{itemize}
        \item $\ell \in [2, 5000]$: TT,TE,EE with $N_{\ell} = 0$ for all spectra
    \end{itemize}
\end{itemize}

\subsection{Results} \label{sec:Results}

Table~\ref{tab:forecast_results} summarizes our Fisher forecast results. As a validation, we check that the Fisher and MCMC pipelines produce bounds that agree within $20\%$ across all configurations (see Table \ref{tab:forecast_results}). Given the greater robustness of MC sampling, we report the MC results for all further analysis. 

Relative to the Planck-only constraint in Table~\ref{tab:dm_ann_limits} ($3.9\times 10^{-28}\,\mathrm{cm}^3\,\mathrm{s}^{-1}\,\mathrm{GeV}^{-1}$),
adding a high-resolution ground-based experiment (CMB-S4–like) tightens the $2\sigma$ upper
limit to $2.13\times10^{-28} \,\mathrm{cm}^3\,\mathrm{s}^{-1}\,\mathrm{GeV}^{-1}$, which corresponds to a $\simeq 46\%$ improvement. Including LiteBIRD data
yields $1.30\times10^{-28}\,\mathrm{cm}^3\,\mathrm{s}^{-1}\,\mathrm{GeV}^{-1}$, which gives an additional $\simeq 38\%$ improvement beyond
CMB-S4$+$Planck and $\simeq 67\%$ relative to Planck alone. A Simons$+$LiteBIRD
configuration yields $1.27\times10^{-28}\,\mathrm{cm}^3\,\mathrm{s}^{-1}\,\mathrm{GeV}^{-1}$, which lies within $\sim 3\%$ of the CMB-S4$+$LiteBIRD bound, indicating that once large-scale polarization is near cosmic variance, further high-$\ell$ noise reduction brings only marginal gains. The Simons$+$LiteBIRD combination effectively saturates (to within $\sim 4\%$) the constraint of a CVL experiment with equivalent sky coverage ($f_{\mathrm{sky}}=0.7$).

Figure~\ref{fig:lmin_exps} shows how the bound varies with the minimum multipole $\ell_{\min}$ included in the analysis. The constraint improves most rapidly as progressively more intermediate multipoles are added, reflecting the fact that these scales carry significant sensitivity to energy injection. For $\ell_{\min} < 200$, we also illustrate the effect of replacing Planck low-$\ell$ noise with LiteBIRD’s higher-sensitivity measurements: doing so yields a noticeable tightening of the limit beyond what is possible with Planck low-$\ell$ noise, demonstrating the additional constraining power that comes from improved large-scale polarization.
These trends are consistent with previous analyses that identify large-scale polarization as a key limiting factor for energy injection constraints~\cite{Green:2018pmd, Padmanabhan:2005es, Madhavacheril:2013cna}.

In Figure~\ref{fig:mass_constraints}, we map the constraints on $p_{\rm ann}$ into the $m_{\chi}$--$\langle \sigma v \rangle$ plane using the appropriate $f_{\rm eff}$ values for three representative annihilation channels and compare them to existing astrophysical bounds.

The resulting projections show that, for hadronic and gauge-boson ($b\bar{b}$ and $W^+W^-$) channels, the improvement provided by future experiments will not provide a more stringent bound with respect to the astronomical ones (based on the non-observation of excess gamma rays from dwarf galaxies and the Galactic center).

If the $\mu^+\mu^-$ channel is considered, the constraining power of future experiments will achieve the most stringent bound up to $\simeq 10$ TeV. However, in any annihilation channel under consideration, the {\it thermal WIMP} paradigm will not be probed at masses larger than $\simeq 30$ GeV, contrary to some previous claims.

\begin{table}[h]
  \caption{\label{tab:forecast_results}Forecasted $2\sigma$ upper limits on $p_{\mathrm{ann}}$, comparing two forecasting methods: MCMC (see \ref{sec:mcmc}), and Fisher forecasting (see \ref{sec:fisher}).} 
    \begin{ruledtabular}
      \begin{tabular}{l S[table-format=1.2] S[table-format=1.2]}
        \textrm{Experiment} &
        \multicolumn{2}{c}{\textrm{$2\sigma$ upper limit [$10^{-28}\,\mathrm{cm^{3}\,s^{-1}\,GeV^{-1}}$]}}\\
        & {MCMC} & {Fisher} \\
        \colrule
        CMB\textendash S4 + Planck   & 2.13 & 2.11 \\
        CMB\textendash S4 + LiteBIRD & 1.30 & 1.52 \\
        SO + LiteBIRD                & 1.27  & 1.49 \\
        CVL                          & 1.23 & 1.25 \\
      \end{tabular}
    \end{ruledtabular}

\end{table}

\section{Effect of Tensor Modes} \label{sec: Effect of Tensor Modes}


Future CMB surveys will measure the B-mode power spectrum with far greater precision than previous datasets, motivating the question of whether BB contributes meaningful additional information to DM annihilation constraints. In the absence of primordial tensor modes ($r=0$), all B-mode power arises from gravitational lensing, which converts E-modes into B-modes. Any modification to the large-scale E-mode signal induced by energy injection is therefore partially remapped into B. If primordial tensors are present, they add BB power on the same large angular scales ($\ell \lesssim 10$) that carry most of the annihilation leverage.

We assess the impact of including BB through a Fisher analysis, in which we consider only two-point CMB power spectra and do not include a separate Fisher contribution from lensing reconstruction (e.g., the $\phi\phi$ power spectrum). As a result, any improvement from adding BB reflects the incremental lensing information contained in the B-mode power spectrum within a two-point, Gaussian approximation.

Within this framework, we extend the covariance matrix in Eq.~\ref{eq:fisher_matrix_base} to include a diagonal B-mode block~\cite{Scott:2016fad, Kasanda:2014fxa}:
\begin{align}
\Sigma_\ell^{(TEB)} &= \frac{2}{(2\ell+1)f_{\mathrm{sky}}} \nonumber\\
&\times
\begin{pmatrix}
C_\ell^{TT}+N_\ell^{TT} & C_\ell^{TE} & 0 \\
C_\ell^{TE} & C_\ell^{EE}+N_\ell^{EE} & 0 \\
0 & 0 & C_\ell^{BB}+N_\ell^{BB}
\end{pmatrix},
\end{align}

where $C_\ell^{BB}$ is the lensed BB spectrum. We assume no delensing,  equal polarization noise ($N_\ell^{BB}\equiv N_\ell^{EE}$), and vanishing TB and EB correlations \cite{Kasanda:2014fxa}. 

We quantify the incremental value of BB by comparing Fisher constraints using 
$\mathrm{TT,TE,EE}$ versus $\mathrm{TT,TE,EE,BB}$ for the same experiment and $\ell$-range, in the following configurations:

\paragraph{CVL.}
For a CVL survey with $f_{\mathrm{sky}}=0.7$ and $\ell_{\max}=3000$,
adding $BB$ tightens the $2\sigma$ upper limit on $p_{\text{ann}}$ by $5.5\%$ relative to the $\mathrm{TT,TE,EE}$ baseline.

\paragraph{LiteBIRD.}
For a more realistic experiment, we consider a LiteBIRD-only experiment as described in Table \ref{tab:forecast_specs}. Including BB tightens the upper limit by $1.5\%$; the minor gain is concentrated at the very largest scales ($\ell <$ 10), where the leverage there is already dominated by E-modes.

We also repeat these forecasts with fixed primordial tensors, $r=10^{-3}$. Including BB tightens the $p_{\rm ann}$ bounds by $5.5\%$ for a CVL survey and $1.7\%$ for LiteBIRD---numerically similar to the $r=0$ case---indicating that small tensor contributions do not enhance sensitivity to $p_{\rm ann}$. The BB response is dominated by gravitational lensing of the E-modes: energy injection alters EE (via an increased free–electron fraction and a broadened visibility function), and lensing remaps those changes into B-modes. Consequently, BB carries little independent information once EE and TE are included, yielding at most marginal additional constraining power; the sensitivity is effectively captured by large-scale E-mode polarization~\cite{Scott:2016fad}.

For an idealized LiteBIRD-like experiment, the incremental leverage from BB remains at the percent level, with any gain concentrated on scales $\ell\lesssim 10$ where EE already dominates. Realistic delensing and noise modeling would further reduce the BB contribution. We therefore view BB primarily as a useful internal consistency check on the polarization signature of energy injection, rather than a significant contribution to the $p_{\rm ann}$ constraint.

\section{Discussion and Conclusions}\label{sec:Discussion}

Our analysis isolates the multipole ranges and experimental features that most directly govern CMB sensitivity to annihilating dark matter. Two main conclusions emerge. First, present constraints are already limited primarily by large–scale $E$-mode polarization; adding high-$\ell$ temperature and polarization from ground-based experiments (ACT/SPT or a CMB-S4–like survey) provides only modest gains once the low-$\ell$ polarization uncertainty is fixed. Second, forecasts that include a space-based mission with precise low-$\ell$ measurements (LiteBIRD) approach the cosmic-variance limit for $f_{\mathrm{sky}}\simeq 0.7$, with only marginal improvement from further reducing high-$\ell$ noise. Taken together, these trends reinforce the picture that the injected-energy signal acts mainly as an effective additional optical depth, enhancing large–scale polarization and thereby making low-$\ell$ sensitivity—not high-$\ell$ resolution—the primary driver of constraining power.

Near-term progress hinges on improved large-scale polarization: absolute calibration, polarization-angle systematics, $1/f$ suppression, and robust foreground control. A LiteBIRD-like survey paired with current or next-generation ground data effectively saturates the information available to the CMB for annihilation histories of the type modeled here. The CMB remains a uniquely clean, model-agnostic bound on annihilation, with forthcoming polarization measurements poised to close the gap to the cosmic-variance floor.

\acknowledgments
We would like to thank Junsong Cang and An\v{z}e Slosar for their helpful clarifications regarding their previous work. We are also grateful to Giovanni Signorelli for his comments regarding LiteBIRD sensitivity. 

DA acknowledges support from the Generalitat Valenciana grant CIGRIS/2021/054, and grant PID2023-151418NB-I00, which is funded by MCIU/AEI/10.13039/501100011033/ FEDER, UE. 

DG acknowledges support from the projects {\it Theoretical Astroparticle Physics (TAsP)} and {\it TEONGRAV} funded by INFN.

CM acknowledges support from the MIT International Science and Technology Initiatives (MISTI), which funded her internship at the University of Pisa. 

This work made use of {\tt CLASS}~\cite{Blas:2011rf}, {\tt Cobaya}~\cite{Torrado:2020dgo,Cobaya_ASCL:2019}, {\tt getdist}~\cite{Lewis:2019xzd}, {\tt matplotlib}~\cite{Hunter:2007}, {\tt NumPy}~\cite{Harris:2020xlr}, {\tt SciPy}~\cite{Virtanen:2019joe}, Webplotdigitizer~\cite{WebPlotDigitizer}, and any dependencies.

\appendix

\section{Prior ranges}\label{app:priors}

\renewcommand{\arraystretch}{1.3} 
\begin{table}[h]
  \caption{Flat prior ranges assumed for the cosmological parameters in our analysis.}
  \begin{ruledtabular}
    \begin{tabular}{l l}
      \textrm{Parameter} & \textrm{Prior} \\
      \colrule
      $p_{\mathrm{ann}}$ [cm$^{3}$\,s$^{-1}$\,GeV$^{-1}$] & [0, 3$\times 10^{-27}$] \\
      $n_s$              & [0.9, 1.1] \\
      $\tau_{\mathrm{reio}}$ & [0, 0.1] \\
      $\log(10^{10} A_s)$ & [2.6, 3.5] \\
      $\Omega_b h^2$     & [0.017, 0.027] \\
      $\Omega_c h^2$     & [0.09, 0.15] \\
      $H_0$ [km/s/Mpc]   & [60, 80] \\
    \end{tabular}
  \end{ruledtabular}
  \label{tab:priors}
\end{table}

In Table~\ref{tab:priors} we give the prior ranges assumed for the cosmological parameters in our analysis.

\section{Triangle plot}\label{app:triangle}

\begin{figure*}
\includegraphics[width=0.9\linewidth]{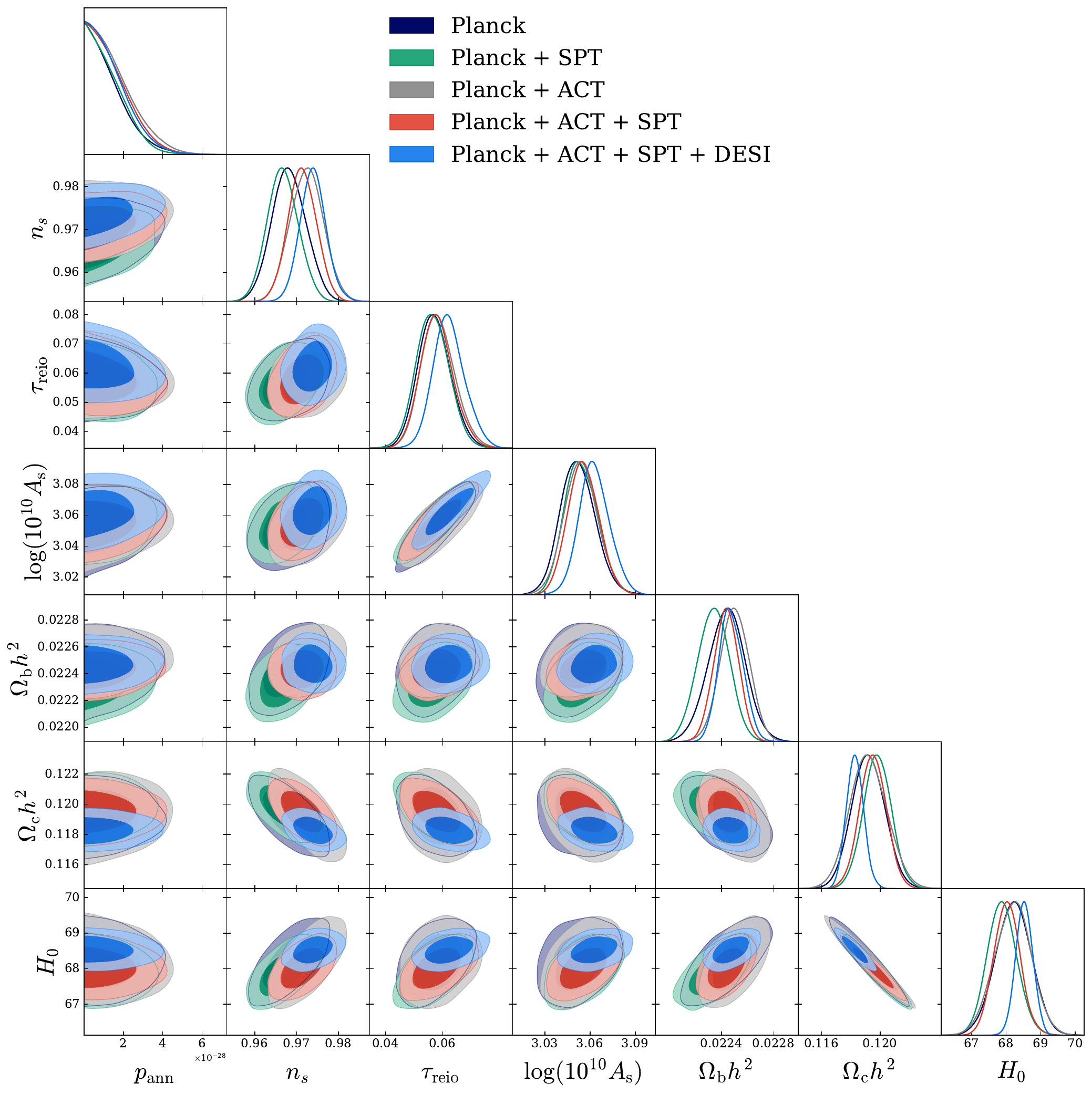}
\caption{
Joint posteriors for
$\{p_{\rm ann},\,n_s,\,\tau_{\rm reio},\,\log(10^{10}A_s),\,\Omega_bh^2,\,\Omega_ch^2,\,H_0\}$
from updated data combinations (see Table \ref{tab:dm_ann_limits}). Shaded regions denote 68\% and 95\% credible intervals; diagonal
panels show the corresponding 1D marginals.
}
\label{fig:full_triangle_plot}
\end{figure*}

In Fig.~\ref{fig:full_triangle_plot} we show the joint posterior distributions for $p_{\mathrm{ann}}$ and the standard six $\Lambda$CDM parameters obtained from various combinations of CMB datasets in Section \ref{sec:data_results}. The addition of high-resolution ground-based measurements (ACT and SPT) to Planck data slightly tightens the constraints on all parameters, most notably $p_{\mathrm{ann}}$. Cross-correlations between $p_{\mathrm{ann}}$ and the recombination parameters $\tau_{\mathrm{reio}}$, $n_s$, and $\log(10^{10}A_s)$ are significantly reduced once ACT+SPT data are included, reflecting the improved polarization sensitivity at both large and small scales. The parameters $\Omega_b h^2$ and $\Omega_c h^2$ remain almost stable across data combinations, while $H_0$ shows only minor shifts, consistent with expectations that energy-injection constraints primarily affect the ionization and primordial-spectrum sectors.

\bibliographystyle{apsrev4-2}
\bibliography{bib_file_inspire}

@inproceedings{2024IAUGA..32P2915R_sne,
    adsurl = {https://ui.adsabs.harvard.edu/abs/2024IAUGA..32P2915R},
    author = {{Rueda-Blanco}, Sebasti{\'a}n and {Delgado-Correal}, Camilo and {Higuera-G.}, Mario-A. and {Torres Arzayus}, Sergio},
    booktitle = {32nd General Assembly International Union (IAUGA 2024)},
    eid = {2915},
    month = {August},
    pages = {2915},
    title = {{Constraining Late-Time Physics to Solve the Hubble Tension}},
    year = {2024}
}

@article{Chen:2003gz,
    author = "Chen, Xue-Lei and Kamionkowski, Marc",
    title = "{Particle decays during the cosmic dark ages}",
    eprint = "astro-ph/0310473",
    archivePrefix = "arXiv",
    doi = "10.1103/PhysRevD.70.043502",
    journal = "Phys. Rev. D",
    volume = "70",
    pages = "043502",
    year = "2004"
}

@article{Peebles:2000pn,
    author = "Peebles, P. J. E. and Seager, Sara and Hu, Wayne",
    title = "{Delayed recombination}",
    eprint = "astro-ph/0004389",
    archivePrefix = "arXiv",
    doi = "10.1086/312831",
    journal = "Astrophys. J. Lett.",
    volume = "539",
    pages = "L1--L4",
    year = "2000"
}

@article{Bean:2003fb,
  title = {Recombining WMAP: Constraints on ionizing and resonance radiation at recombination},
  author = {Bean, Rachel and Melchiorri, Alessandro and Silk, Joe},
  journal = {Phys. Rev. D},
  volume = {68},
  issue = {8},
  pages = {083501},
  numpages = {6},
  year = {2003},
  month = {Oct},
  publisher = {American Physical Society},
  doi = {10.1103/PhysRevD.68.083501},
  url = {https://link.aps.org/doi/10.1103/PhysRevD.68.083501}
}

@article{Hansen:2003yj,
    author = "Hansen, Steen H. and Haiman, Zoltan",
    title = "{Do we need stars to reionize the universe at high redshifts? Early reionization by decaying heavy sterile neutrinos}",
    eprint = "astro-ph/0305126",
    archivePrefix = "arXiv",
    doi = "10.1086/379636",
    journal = "Astrophys. J.",
    volume = "600",
    pages = "26--31",
    year = "2004"
}

@article{Steigman:2012nb,
    author = "Steigman, Gary and Dasgupta, Basudeb and Beacom, John F.",
    title = "{Precise Relic WIMP Abundance and its Impact on Searches for Dark Matter Annihilation}",
    eprint = "1204.3622",
    archivePrefix = "arXiv",
    primaryClass = "hep-ph",
    doi = "10.1103/PhysRevD.86.023506",
    journal = "Phys. Rev. D",
    volume = "86",
    pages = "023506",
    year = "2012"
}

@article{Pierpaoli:2003rz,
    author = "Pierpaoli, Elena",
    title = "{Decaying particles and the reionization history of the universe}",
    eprint = "astro-ph/0310375",
    archivePrefix = "arXiv",
    doi = "10.1103/PhysRevLett.92.031301",
    journal = "Phys. Rev. Lett.",
    volume = "92",
    pages = "031301",
    year = "2004"
}

@article{Doroshkevich:2002wy,
    author = "Doroshkevich, A. G. and Naselsky, I. P. and Naselsky, P. D. and Novikov, I. D.",
    title = "{Ionization history of the cosmic plasma in the light of the recent cbi and future planck data}",
    eprint = "astro-ph/0208114",
    archivePrefix = "arXiv",
    doi = "10.1086/367819",
    journal = "Astrophys. J.",
    volume = "586",
    pages = "709--717",
    year = "2003"
}

@article{ACT:2025xdm,
    author = "Naess, Sigurd and others",
    collaboration = "Atacama Cosmology Telescope",
    title = "{The Atacama Cosmology Telescope: DR6 maps}",
    eprint = "2503.14451",
    archivePrefix = "arXiv",
    primaryClass = "astro-ph.CO",
    reportNumber = "FERMILAB-PUB-25-0160-PPD",
    doi = "10.1088/1475-7516/2025/11/061",
    journal = "JCAP",
    volume = "11",
    pages = "061",
    year = "2025"
}

@article{ACT:2025fju,
    author = "Louis, Thibaut and others",
    collaboration = "Atacama Cosmology Telescope",
    title = "{The Atacama Cosmology Telescope: DR6 power spectra, likelihoods and {\ensuremath{\Lambda}}CDM parameters}",
    eprint = "2503.14452",
    archivePrefix = "arXiv",
    primaryClass = "astro-ph.CO",
    reportNumber = "FERMILAB-PUB-25-0071-PPD",
    doi = "10.1088/1475-7516/2025/11/062",
    journal = "JCAP",
    volume = "11",
    pages = "062",
    year = "2025"
}

@article{ACT:2025tim,
    author = "Calabrese, Erminia and others",
    collaboration = "Atacama Cosmology Telescope",
    title = "{The Atacama Cosmology Telescope: DR6 constraints on extended cosmological models}",
    eprint = "2503.14454",
    archivePrefix = "arXiv",
    primaryClass = "astro-ph.CO",
    reportNumber = "FERMILAB-PUB-25-0157-PPD",
    doi = "10.1088/1475-7516/2025/11/063",
    journal = "JCAP",
    volume = "11",
    pages = "063",
    year = "2025"
}

@article{Adams:1998nr,
    archiveprefix = {arXiv},
    author = {Adams, Jennifer A. and Sarkar, Subir and Sciama, D. W.},
    doi = {10.1046/j.1365-8711.1998.02017.x},
    eprint = {astro-ph/9805108},
    journal = {Mon. Not. Roy. Astron. Soc.},
    pages = {210--214},
    reportnumber = {UUITP-1-98, OUTP-98-24-P, SISSA-37-98-A},
    title = {{CMB anisotropy in the decaying neutrino cosmology}},
    volume = {301},
    year = {1998}
}

@article{Arcadi:2017kky,
    archiveprefix = {arXiv},
    author = {Arcadi, Giorgio and Dutra, Ma{\'\i}ra and Ghosh, Pradipta and Lindner, Manfred and Mambrini, Yann and Pierre, Mathias and Profumo, Stefano and Queiroz, Farinaldo S.},
    doi = {10.1140/epjc/s10052-018-5662-y},
    eprint = {1703.07364},
    journal = {Eur. Phys. J. C},
    number = {3},
    pages = {203},
    primaryclass = {hep-ph},
    title = {{The waning of the WIMP? A review of models, searches, and constraints}},
    volume = {78},
    year = {2018}
}

@article{Blas:2011rf,
    author = {Diego Blas and Julien Lesgourgues and Thomas Tram},
    doi = {10.1088/1475-7516/2011/07/034},
    issn = {1475-7516},
    journal = {Journal of Cosmology and Astroparticle Physics},
    month = {July},
    number = {07},
    pages = {034–034},
    publisher = {IOP Publishing},
    title = {The Cosmic Linear Anisotropy Solving System (CLASS).
Part II: Approximation schemes},
    url = {http://dx.doi.org/10.1088/1475-7516/2011/07/034},
    volume = {2011},
    year = {2011}
}

@article{Cang:2020exa,
    author = {Cang, Junsong and Gao, Yu and Ma, Yin-Zhe},
    doi = {10.1103/PhysRevD.102.103005},
    issue = {10},
    journal = {Phys. Rev. D},
    month = {Nov},
    numpages = {10},
    pages = {103005},
    publisher = {American Physical Society},
    title = {Probing dark matter with future CMB measurements},
    url = {https://link.aps.org/doi/10.1103/PhysRevD.102.103005},
    volume = {102},
    year = {2020}
}

@article{Cirelli:2024ssz,
    adsurl = {https://ui.adsabs.harvard.edu/abs/2024arXiv240601705C},
    archiveprefix = {arXiv},
    author = {Cirelli, Marco and Strumia, Alessandro and Zupan, Jure},
    doi = {10.48550/arXiv.2406.01705},
    eid = {arXiv:2406.01705},
    eprint = {2406.01705},
    journal = {arXiv e-prints},
    month = {6},
    pages = {arXiv:2406.01705},
    primaryclass = {hep-ph},
    title = {{Dark Matter}},
    year = {2024}
}

@article{Cline:2013fm,
    archiveprefix = {arXiv},
    author = {Cline, James M. and Scott, Pat},
    doi = {10.1088/1475-7516/2013/03/044},
    eprint = {1301.5908},
    journal = {JCAP},
    note = {[Erratum: JCAP 05, E01 (2013)]},
    pages = {044},
    primaryclass = {astro-ph.CO},
    title = {{Dark Matter CMB Constraints and Likelihoods for Poor Particle Physicists}},
    volume = {03},
    year = {2013}
}

@misc{CMBS4Web,
    author = {{CMB-S4 Collaboration}},
    howpublished = {\url{https://cmb-s4.org/}},
    note = {Accessed: 2025-09-01},
    title = {{CMB-S4}: Cosmic Microwave Background Stage-4 (official website)}
}

@book{CMB-S4:2016ple,
    author = "Abazajian, Kevork N. and others",
    collaboration = "CMB-S4",
    title = "{CMB-S4 Science Book, First Edition}",
    eprint = "1610.02743",
    archivePrefix = "arXiv",
    primaryClass = "astro-ph.CO",
    reportNumber = "FERMILAB-FN-1024-A-AE",
    doi = "10.2172/1352047",
    month = "10",
    year = "2016"
}

@article{Diamanti:2013bia,
    archiveprefix = {arXiv},
    author = {Diamanti, Roberta and Lopez-Honorez, Laura and Mena, Olga and Palomares-Ruiz, Sergio and Vincent, Aaron C.},
    doi = {10.1088/1475-7516/2014/02/017},
    eprint = {1308.2578},
    journal = {JCAP},
    pages = {017},
    primaryclass = {astro-ph.CO},
    reportnumber = {IFIC-13-54},
    title = {{Constraining Dark Matter Late-Time Energy Injection: Decays and P-Wave Annihilations}},
    volume = {02},
    year = {2014}
}

@article{Euclid:2021qvm,
    author = "Ili{\'c}, S. and others",
    collaboration = "Euclid",
    title = "{Euclid preparation. XV. Forecasting cosmological constraints for the Euclid and CMB joint analysis}",
    eprint = "2106.08346",
    archivePrefix = "arXiv",
    primaryClass = "astro-ph.CO",
    doi = "10.1051/0004-6361/202141556",
    journal = "Astron. Astrophys.",
    volume = "657",
    pages = "A91",
    year = "2022"
}

@article{Finkbeiner:2011dx,
    archiveprefix = {arXiv},
    author = {Finkbeiner, Douglas P. and Galli, Silvia and Lin, Tongyan and Slatyer, Tracy R.},
    doi = {10.1103/PhysRevD.85.043522},
    eprint = {1109.6322},
    journal = {Phys. Rev. D},
    pages = {043522},
    primaryclass = {astro-ph.CO},
    title = {{Searching for Dark Matter in the CMB: A Compact Parameterization of Energy Injection from New Physics}},
    volume = {85},
    year = {2012}
}

@article{Galli:2009zc,
    archiveprefix = {arXiv},
    author = {Galli, Silvia and Iocco, Fabio and Bertone, Gianfranco and Melchiorri, Alessandro},
    doi = {10.1103/PhysRevD.80.023505},
    eprint = {0905.0003},
    journal = {Phys. Rev. D},
    pages = {023505},
    primaryclass = {astro-ph.CO},
    reportnumber = {SACLAY-T09-046},
    title = {{CMB constraints on Dark Matter models with large annihilation cross-section}},
    volume = {80},
    year = {2009}
}

@article{Galli:2011rz,
    archiveprefix = {arXiv},
    author = {Galli, Silvia and Iocco, Fabio and Bertone, Gianfranco and Melchiorri, Alessandro},
    doi = {10.1103/PhysRevD.84.027302},
    eprint = {1106.1528},
    journal = {Phys. Rev. D},
    pages = {027302},
    primaryclass = {astro-ph.CO},
    title = {{Updated CMB constraints on Dark Matter annihilation cross-sections}},
    volume = {84},
    year = {2011}
}

@article{Galli:2013dna,
    archiveprefix = {arXiv},
    author = {Galli, Silvia and Slatyer, Tracy R. and Valdes, Marcos and Iocco, Fabio},
    doi = {10.1103/PhysRevD.88.063502},
    eprint = {1306.0563},
    journal = {Phys. Rev. D},
    pages = {063502},
    primaryclass = {astro-ph.CO},
    title = {{Systematic Uncertainties In Constraining Dark Matter Annihilation From The Cosmic Microwave Background}},
    volume = {88},
    year = {2013}
}

@article{Giesen:2012rp,
    archiveprefix = {arXiv},
    author = {Giesen, Gaelle and Lesgourgues, Julien and Audren, Benjamin and Ali-Haimoud, Yacine},
    doi = {10.1088/1475-7516/2012/12/008},
    eprint = {1209.0247},
    journal = {JCAP},
    pages = {008},
    primaryclass = {astro-ph.CO},
    reportnumber = {CERN-PH-TH-2012-216, LAPTH-038-12},
    title = {{CMB photons shedding light on dark matter}},
    volume = {12},
    year = {2012}
}

@article{Green:2018pmd,
    author = {Green, Daniel and Meerburg, P. Daniel and Meyers, Joel},
    doi = {10.1088/1475-7516/2019/04/025},
    journal = {Journal of Cosmology and Astroparticle Physics},
    month = {apr},
    number = {04},
    pages = {025},
    publisher = {},
    title = {Aspects of dark matter annihilation in cosmology},
    url = {https://dx.doi.org/10.1088/1475-7516/2019/04/025},
    volume = {2019},
    year = {2019}
}

@article{Hutsi:2011vx,
    archiveprefix = {arXiv},
    author = {Hutsi, Gert and Chluba, Jens and Hektor, Andi and Raidal, Martti},
    doi = {10.1051/0004-6361/201116914},
    eprint = {1103.2766},
    journal = {Astron. Astrophys.},
    pages = {A26},
    primaryclass = {astro-ph.CO},
    title = {{WMAP7 and future CMB constraints on annihilating dark matter: implications on GeV-scale WIMPs}},
    volume = {535},
    year = {2011}
}

@article{Kanzaki:2009hf,
    archiveprefix = {arXiv},
    author = {Kanzaki, Toru and Kawasaki, Masahiro and Nakayama, Kazunori},
    doi = {10.1143/PTP.123.853},
    eprint = {0907.3985},
    journal = {Prog. Theor. Phys.},
    pages = {853--865},
    primaryclass = {astro-ph.CO},
    reportnumber = {ICRR-REPORT-550, IPMU-09-0087},
    title = {{Effects of Dark Matter Annihilation on the Cosmic Microwave Background}},
    volume = {123},
    year = {2010}
}

@article{Lewis:2019xzd,
   author = "Lewis, Antony",
   title = "{GetDist: a Python package for analysing Monte Carlo samples}",
   eprint = "1910.13970",
   archivePrefix = "arXiv",
   primaryClass = "astro-ph.IM",
   doi = "10.1088/1475-7516/2025/08/025",
   journal = "JCAP",
   volume = "08",
   pages = "025",
   year = "2025"
}

@article{Harris:2020xlr,
    author = "Harris, Charles R. and others",
    title = "{Array programming with NumPy}",
    eprint = "2006.10256",
    archivePrefix = "arXiv",
    primaryClass = "cs.MS",
    doi = "10.1038/s41586-020-2649-2",
    journal = "Nature",
    volume = "585",
    number = "7825",
    pages = "357--362",
    year = "2020"
}

@article{Virtanen:2019joe,
    author = "Virtanen, Pauli and others",
    title = "{SciPy 1.0--Fundamental algorithms for scientific computing in Python}",
    eprint = "1907.10121",
    archivePrefix = "arXiv",
    primaryClass = "cs.MS",
    doi = "10.1038/s41592-019-0686-2",
    journal = "Nature Meth.",
    volume = "17",
    pages = "261",
    year = "2020"
}

@misc{Cobaya_ASCL:2019,
    author       = {{Torrado}, Jes{\'u}s and {Lewis}, Antony},
    title        = "{Cobaya: Bayesian analysis in cosmology}",
    howpublished = {Astrophysics Source Code Library, record ascl:1910.019},
    year         = 2019,
    month        = oct,
    eid          = {ascl:1910.019},
    note         = {ascl:1910.019},
    url          = {https://ui.adsabs.harvard.edu/abs/2019ascl.soft10019T}
}

@misc{WebPlotDigitizer,
    author       = {Ankit Rohatgi},
    title        = {WebPlotDigitizer},
    howpublished = {\url{https://automeris.io}},
    note         = {Version 5.2}
}

@Article{Hunter:2007,
  Author    = {Hunter, J. D.},
  Title     = {Matplotlib: A 2D graphics environment},
  Journal   = {Computing in Science \& Engineering},
  Volume    = {9},
  Number    = {3},
  Pages     = {90--95},
  publisher = {IEEE COMPUTER SOC},
  doi       = {10.1109/MCSE.2007.55},
  year      = 2007
}

@article{Kasanda:2014fxa,
    author = {Kasanda, Simon Muya and Moodley, Kavilan},
    doi = {10.1088/1475-7516/2014/12/041},
    issn = {1475-7516},
    journal = {Journal of Cosmology and Astroparticle Physics},
    month = {December},
    number = {12},
    pages = {041–041},
    publisher = {IOP Publishing},
    title = {CMB lensing forecasts for constraining the primordial perturbations: adding to the CMB temperature and polarization information},
    url = {http://dx.doi.org/10.1088/1475-7516/2014/12/041},
    volume = {2014},
    year = {2014}
}

@article{Kawasaki:2021etm,
    archiveprefix = {arXiv},
    author = {Kawasaki, Masahiro and Nakatsuka, Hiromasa and Nakayama, Kazunori and Sekiguchi, Toyokazu},
    doi = {10.1088/1475-7516/2021/12/015},
    eprint = {2105.08334},
    journal = {JCAP},
    number = {12},
    pages = {015},
    primaryclass = {astro-ph.CO},
    title = {{Revisiting CMB constraints on dark matter annihilation}},
    volume = {12},
    year = {2021}
}

@article{Knox:1995dq,
    author = {Knox, Lloyd},
    doi = {10.1103/PhysRevD.52.4307},
    issue = {8},
    journal = {Phys. Rev. D},
    month = {Oct},
    numpages = {0},
    pages = {4307--4318},
    publisher = {American Physical Society},
    title = {Determination of inflationary observables by cosmic microwave background anisotropy experiments},
    url = {https://link.aps.org/doi/10.1103/PhysRevD.52.4307},
    volume = {52},
    year = {1995}
}

@article{Calabrese:2016eii,
    author = "Calabrese, Erminia and Alonso, David and Dunkley, Jo",
    title = "{Complementing the ground-based CMB-S4 experiment on large scales with the PIXIE satellite}",
    eprint = "1611.10269",
    archivePrefix = "arXiv",
    primaryClass = "astro-ph.CO",
    doi = "10.1103/PhysRevD.95.063504",
    journal = "Phys. Rev. D",
    volume = "95",
    number = "6",
    pages = "063504",
    year = "2017"
}

@article{Galli:2010it,
    author = "Galli, Silvia and Martinelli, Matteo and Melchiorri, Alessandro and Pagano, Luca and Sherwin, Blake D. and Spergel, David N.",
    title = "{Constraining Fundamental Physics with Future CMB Experiments}",
    eprint = "1005.3808",
    archivePrefix = "arXiv",
    primaryClass = "astro-ph.CO",
    doi = "10.1103/PhysRevD.82.123504",
    journal = "Phys. Rev. D",
    volume = "82",
    pages = "123504",
    year = "2010"
}

@article{Galli:2014kla,
    author = "Galli, Silvia and Benabed, Karim and Bouchet, Fran{\c{c}}ois and Cardoso, Jean-Fran{\c{c}}ois and Elsner, Franz and Hivon, Eric and Mangilli, Anna and Prunet, Simon and Wandelt, Benjamin",
    title = "{CMB Polarization can constrain cosmology better than CMB temperature}",
    eprint = "1403.5271",
    archivePrefix = "arXiv",
    primaryClass = "astro-ph.CO",
    doi = "10.1103/PhysRevD.90.063504",
    journal = "Phys. Rev. D",
    volume = "90",
    number = "6",
    pages = "063504",
    year = "2014"
}

@article{LiteBIRD:2022cnt,
    author = "Allys, E. and others",
    collaboration = "LiteBIRD",
    title = "{Probing Cosmic Inflation with the LiteBIRD Cosmic Microwave Background Polarization Survey}",
    eprint = "2202.02773",
    archivePrefix = "arXiv",
    primaryClass = "astro-ph.IM",
    doi = "10.1093/ptep/ptac150",
    journal = "PTEP",
    volume = "2023",
    number = "4",
    pages = "042F01",
    year = "2023"
}

@article{Lopez-Honorez:2013cua,
    archiveprefix = {arXiv},
    author = {Lopez-Honorez, Laura and Mena, Olga and Palomares-Ruiz, Sergio and Vincent, Aaron C.},
    doi = {10.1088/1475-7516/2013/07/046},
    eprint = {1303.5094},
    journal = {JCAP},
    pages = {046},
    primaryclass = {astro-ph.CO},
    reportnumber = {IFIC-13-016, CFTP-13-007},
    title = {{Constraints on dark matter annihilation from CMB observations before Planck}},
    volume = {07},
    year = {2013}
}

@article{Madhavacheril:2013cna,
    author = {Madhavacheril, Mathew S. and Sehgal, Neelima and Slatyer, Tracy R.},
    doi = {10.1103/physrevd.89.103508},
    issn = {1550-2368},
    journal = {Physical Review D},
    month = {May},
    number = {10},
    publisher = {American Physical Society (APS)},
    title = {Current dark matter annihilation constraints from CMB and low-redshift data},
    url = {http://dx.doi.org/10.1103/PhysRevD.89.103508},
    volume = {89},
    year = {2014}
}

@article{Micheli:2024hfe,
    archiveprefix = {arXiv},
    author = {Micheli, Silvia and others},
    doi = {10.1117/12.3018553},
    eprint = {2407.15294},
    journal = {Proc. SPIE Int. Soc. Opt. Eng.},
    pages = {131022R},
    primaryclass = {astro-ph.IM},
    title = {{Systematic effects induced by half-wave plate differential optical load and TES nonlinearity for LiteBIRD}},
    url = {https://arxiv.org/abs/2407.15294},
    volume = {13102},
    year = {2024}
}

@article{Natarajan:2012ry,
    author = {Natarajan, Aravind},
    doi = {10.1103/physrevd.85.083517},
    issn = {1550-2368},
    journal = {Physical Review D},
    month = {April},
    number = {8},
    publisher = {American Physical Society (APS)},
    title = {Closer look at CMB constraints on WIMP dark matter},
    url = {http://dx.doi.org/10.1103/PhysRevD.85.083517},
    volume = {85},
    year = {2012}
}

@article{Padmanabhan:2005es,
    author = {Padmanabhan, Nikhil and Finkbeiner, Douglas P.},
    doi = {10.1103/PhysRevD.72.023508},
    issue = {2},
    journal = {Phys. Rev. D},
    month = {Jul},
    numpages = {13},
    pages = {023508},
    publisher = {American Physical Society},
    title = {Detecting dark matter annihilation with CMB polarization: Signatures and experimental prospects},
    url = {https://link.aps.org/doi/10.1103/PhysRevD.72.023508},
    volume = {72},
    year = {2005}
}

@article{Planck:2015fie,
    archiveprefix = {arXiv},
    author = {Ade, P. A. R. and others},
    collaboration = {Planck},
    doi = {10.1051/0004-6361/201525830},
    eprint = {1502.01589},
    journal = {Astron. Astrophys.},
    pages = {A13},
    primaryclass = {astro-ph.CO},
    title = {{Planck 2015 results. XIII. Cosmological parameters}},
    volume = {594},
    year = {2016}
}

@article{gelman_inference_1992,
	title = {Inference from {Iterative} {Simulation} {Using} {Multiple} {Sequences}},
	volume = {7},
	issn = {0883-4237},
	url = {https://projecteuclid.org/journals/statistical-science/volume-7/issue-4/Inference-from-Iterative-Simulation-Using-Multiple-Sequences/10.1214/ss/1177011136.full},
	doi = {10.1214/ss/1177011136},
	number = {4},
	urldate = {2023-11-25},
	journal = {Statistical Science},
	author = {Gelman, Andrew and Rubin, Donald B.},
	month = nov,
	year = {1992}
	
}

@article{Planck:2018vyg,
    author = "Aghanim, N. and others",
    collaboration = "Planck",
    title = "{Planck 2018 results. VI. Cosmological parameters}",
    eprint = "1807.06209",
    archivePrefix = "arXiv",
    primaryClass = "astro-ph.CO",
    doi = "10.1051/0004-6361/201833910",
    journal = "Astron. Astrophys.",
    volume = "641",
    pages = "A6",
    year = "2020",
    note = "[Erratum: Astron.Astrophys. 652, C4 (2021)]"
}

@article{Sailer:2021yzm,
    author = {Sailer, Noah and Castorina, Emanuele and Ferraro, Simone and White, Martin},
    doi = {10.1088/1475-7516/2021/12/049},
    issn = {1475-7516},
    journal = {Journal of Cosmology and Astroparticle Physics},
    month = {December},
    number = {12},
    pages = {049},
    publisher = {IOP Publishing},
    title = {Cosmology at high redshift — a probe of fundamental physics},
    url = {http://dx.doi.org/10.1088/1475-7516/2021/12/049},
    volume = {2021},
    year = {2021}
}

@article{Scott:2016fad,
    author = {Scott, Douglas and Contreras, Dagoberto and Narimani, Ali and Ma, Yin-Zhe},
    doi = {10.1088/1475-7516/2016/06/046},
    issn = {1475-7516},
    journal = {Journal of Cosmology and Astroparticle Physics},
    month = {June},
    number = {06},
    pages = {046–046},
    publisher = {IOP Publishing},
    title = {The information content of cosmic microwave background anisotropies},
    url = {http://dx.doi.org/10.1088/1475-7516/2016/06/046},
    volume = {2016},
    year = {2016}
}

@article{SimonsObservatory:2018koc,
    author = "Ade, Peter and others",
    collaboration = "Simons Observatory",
    title = "{The Simons Observatory: Science goals and forecasts}",
    eprint = "1808.07445",
    archivePrefix = "arXiv",
    primaryClass = "astro-ph.CO",
    doi = "10.1088/1475-7516/2019/02/056",
    journal = "JCAP",
    volume = "02",
    pages = "056",
    year = "2019"
}

@article{Slatyer:2009yq,
    archiveprefix = {arXiv},
    author = {Slatyer, Tracy R. and Padmanabhan, Nikhil and Finkbeiner, Douglas P.},
    doi = {10.1103/PhysRevD.80.043526},
    eprint = {0906.1197},
    journal = {Phys. Rev. D},
    pages = {043526},
    primaryclass = {astro-ph.CO},
    title = {{CMB Constraints on WIMP Annihilation: Energy Absorption During the Recombination Epoch}},
    volume = {80},
    year = {2009}
}

@article{Slatyer:2015jla,
    archiveprefix = {arXiv},
    author = {Slatyer, Tracy R.},
    doi = {10.1103/PhysRevD.93.023527},
    eprint = {1506.03811},
    journal = {Phys. Rev. D},
    number = {2},
    pages = {023527},
    primaryclass = {hep-ph},
    reportnumber = {MIT-CTP-4682},
    title = {{Indirect dark matter signatures in the cosmic dark ages. I. Generalizing the bound on s-wave dark matter annihilation from Planck results}},
    volume = {93},
    year = {2016}
}

@article{Slatyer:2015kla,
    archiveprefix = {arXiv},
    author = {Slatyer, Tracy R.},
    doi = {10.1103/PhysRevD.93.023521},
    eprint = {1506.03812},
    journal = {Phys. Rev. D},
    number = {2},
    pages = {023521},
    primaryclass = {astro-ph.CO},
    reportnumber = {MIT-CTP-4683},
    title = {{Indirect Dark Matter Signatures in the Cosmic Dark Ages II. Ionization, Heating and Photon Production from Arbitrary Energy Injections}},
    volume = {93},
    year = {2016}
}

@article{Slatyer:2016qyl,
    archiveprefix = {arXiv},
    author = {Slatyer, Tracy R. and Wu, Chih-Liang},
    doi = {10.1103/PhysRevD.95.023010},
    eprint = {1610.06933},
    journal = {Phys. Rev. D},
    number = {2},
    pages = {023010},
    primaryclass = {astro-ph.CO},
    reportnumber = {MIT-CTP-4842},
    title = {{General Constraints on Dark Matter Decay from the Cosmic Microwave Background}},
    volume = {95},
    year = {2017}
}

@article{SO:2024ntl,
    archiveprefix = {arXiv},
    author = {Galitzki, Nicholas and others},
    collaboration = {SO},
    doi = {10.3847/1538-4365/ad64c9},
    eprint = {2405.05550},
    journal = {Astrophys. J. Suppl.},
    number = {2},
    pages = {33},
    primaryclass = {astro-ph.IM},
    title = {{The Simons Observatory: Design, Integration, and Testing of the Small Aperture Telescopes}},
    url = {https://arxiv.org/abs/2405.05550},
    volume = {274},
    year = {2024}
}

@misc{SPT-3G:2025bzu,
    author = "Camphuis, E. and others",
    collaboration = "SPT-3G",
    title = "{SPT-3G D1: CMB temperature and polarization power spectra and cosmology from 2019 and 2020 observations of the SPT-3G Main field}",
    eprint = "2506.20707",
    archivePrefix = "arXiv",
    primaryClass = "astro-ph.CO",
    reportNumber = "FERMILAB-PUB-25-0144-PPD",
    month = "6",
    year = "2025"
}

@article{Stocker:2018avm,
    author = {Stöcker, Patrick and Krämer, Michael and Lesgourgues, Julien and Poulin, Vivian},
    doi = {10.1088/1475-7516/2018/03/018},
    issn = {1475-7516},
    journal = {Journal of Cosmology and Astroparticle Physics},
    month = {March},
    note = {arXiv:1801.01871 [astro-ph, physics:hep-ph]},
    number = {03},
    pages = {018--018},
    shorttitle = {Exotic energy injection with {ExoCLASS}},
    title = {Exotic energy injection with {ExoCLASS}: {Application} to the {Higgs} portal model and evaporating black holes},
    url = {http://arxiv.org/abs/1801.01871},
    urldate = {2023-04-12},
    volume = {2018},
    year = {2018}
}

@article{Torrado:2020dgo,
    author = "Torrado, Jesus and Lewis, Antony",
    title = "{Cobaya: Code for Bayesian Analysis of hierarchical physical models}",
    eprint = "2005.05290",
    archivePrefix = "arXiv",
    primaryClass = "astro-ph.IM",
    reportNumber = "TTK-20-15",
    doi = "10.1088/1475-7516/2021/05/057",
    journal = "JCAP",
    volume = "05",
    pages = "057",
    year = "2021"
}

@article{Wang:2025tdx,
    archiveprefix = {arXiv},
    author = {Wang, Yu-Ning and Duan, Xin-Chen and Tang, Tian-Peng and Wang, Ziwei and Tsai, Yue-Lin Sming},
    doi = {10.1088/1475-7516/2025/08/059},
    eprint = {2502.18263},
    journal = {JCAP},
    pages = {059},
    primaryclass = {hep-ph},
    title = {{Exploring sub-GeV dark matter via s-wave, p-wave, and resonance annihilation with CMB data}},
    url = {https://arxiv.org/abs/2502.18263},
    volume = {08},
    year = {2025}
}

@article{Weniger:2013hja,
    archiveprefix = {arXiv},
    author = {Weniger, Christoph and Serpico, Pasquale D. and Iocco, Fabio and Bertone, Gianfranco},
    doi = {10.1103/PhysRevD.87.123008},
    eprint = {1303.0942},
    journal = {Phys. Rev. D},
    number = {12},
    pages = {123008},
    primaryclass = {astro-ph.CO},
    title = {{CMB bounds on dark matter annihilation: Nucleon energy-losses after recombination}},
    volume = {87},
    year = {2013}
}

@article{Wu:2014hta,
    author = {Wu, W. L. K. and Errard, J. and Dvorkin, C. and Kuo, C. L. and Lee, A. T. and McDonald, P. and Slosar, A. and Zahn, O.},
    doi = {10.1088/0004-637x/788/2/138},
    issn = {1538-4357},
    journal = {The Astrophysical Journal},
    month = {June},
    number = {2},
    pages = {138},
    publisher = {American Astronomical Society},
    title = {A GUIDE TO DESIGNING FUTURE GROUND-BASED COSMIC MICROWAVE BACKGROUND EXPERIMENTS},
    url = {http://dx.doi.org/10.1088/0004-637X/788/2/138},
    volume = {788},
    year = {2014}
}

@article{Zhang:2023usm,
    author = {Zhang, Zi-Xuan and Wang, Yi-Ming and Cang, Junsong and Zhang, Zirui and Liu, Yang and Li, Si-Yu and Gao, Yu and Li, Hong},
    doi = {10.1088/1475-7516/2023/10/002},
    journal = {Journal of Cosmology and Astroparticle Physics},
    month = {oct},
    number = {10},
    pages = {002},
    publisher = {IOP Publishing},
    title = {Dark matter search with CMB: a study of foregrounds},
    url = {https://dx.doi.org/10.1088/1475-7516/2023/10/002},
    volume = {2023},
    year = {2023}
}
\end{document}